# Deep-Dive Analysis of Selfish and Stubborn Mining in Bitcoin and Ethereum

Runkai Yang, Xiaolin Chang, Jelena Mišić, Vojislav B. Mišić

**Abstract**—Bitcoin and Ethereum are the top two blockchain-based cryptocurrencies whether from cryptocurrency market cap or popularity. However, they are vulnerable to selfish mining and stubborn mining due to that both of them adopt Proof-of-Work consensus mechanism. In this paper, we develop a novel Markov model, which can study selfish mining and seven kinds of stubborn mining in both Bitcoin and Ethereum. The formulas are derived to calculate several key metrics, including relative revenue of miners, blockchain performance in terms of stale block ratio and transactions per second, and blockchain security in terms of resistance against double-spending attacks. Numerical analysis is conducted to investigate the quantitative relationship between the relative-revenue-optimal mining strategy for malicious miners and two miner features in Bitcoin and Ethereum, respectively. The quantitative analysis results can assist honest miners in detecting whether there is any malicious miner in the system and setting the threshold of mining node's hash power in order to prevent malicious miners from making profit through selfish and stubborn mining.

## 1 INTRODUCTION

Bitcoin and Ethereum are now the two largest and most popular blockchain-based cryptocurrencies in the world [1][2]. They both employ the most widely used consensus mechanism, Proof-of-Work (PoW). In a PoW blockchain system, a large number of miners mine blocks by trying to solve a math puzzle and the miner who first solves the puzzle wins the right to add the next block to the blockchain. The power of miners solving puzzles is denoted as hash power. The miner who produces a regular block (a block which is in the main chain) is rewarded with a pre-defined number of tokens which are known as a mining reward. To increase the chance of producing a block and then getting a reward, several miners can form a mining pool and conduct cooperative mining. In a PoW system, the difficulty of mining a block is usually adjusted to alleviate the impact of the varying hash power and other factors on the regular block generation time (10 min in Bitcoin and 13 sec in Ethereum [3]). This adjustment mechanism is denoted as difficulty adjustment.

PoW blockchain faces many security threats, such as double-spending attack [4], eclipses attack [5] and selfish mining attack [6]. Pow Blockchain security incidents have been reported over the past few years. In 2016, attackers managed to attack 'The DAO' and stolen over 50 million USD of Ethers (the native cryptocurrency of Ethereum) [7]. Bitfinex, a cryptocurrency exchange, lost about 120,000 Bitcoins in 2016 [8]. In 2018, five blockchain-based cryptocurrencies lost 5 million USD because of 51% attack [9]. Ethereum Classic suffered selfish mining in 2019 [10].

This paper focuses on two types of malicious mining, namely, selfish and stubborn mining which are detailed in Sec. 3.1. There are two types of miners, honest and malicious miners. The former type always adopts the honest mining strategy to conduct honest mining, namely, always mining in the longest branch and publishing a block immediately after being produced. The latter type forms a Malicious Pool (MP) to get more undeserved mining revenue (the total reward for a mining node, which is an individual miner or a mining pool). Different from honest miners, the malicious pool can conduct honest mining by adopting the honest mining strategy or can conduct malicious mining by a malicious mining strategy, depending on their hash power and other factors. If a block is produced by the MP, it can be published strategically instead of being published immediately.

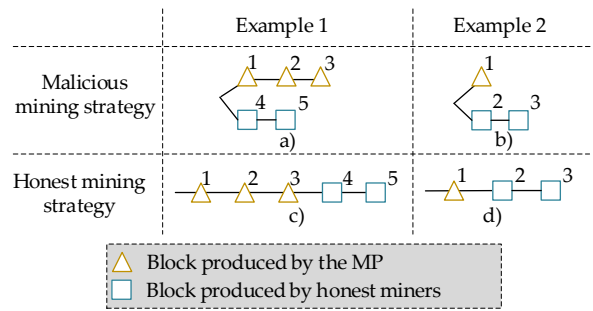

Fig. 1 Two examples of the differences between the honest mining strategy and a malicious mining strategy. The numbers next to the blocks denote the order of the blocks produced.

Malicious mining can lead to hash power waste and then honest miners getting less mining revenue. Fig. 1 gives two examples to illustrate the block chains when the MP employs malicious and honest mining strategies in Bitcoin, respectively. In Example 1 (shown in Fig. 1-a and Fig. 1-c), five blocks are produced in numerical order. Fig. 1-a indicates that under a malicious strategy, the branch with three blocks all produced by the MP is the main chain. Since only blocks on the main chain can bring reward and the reward of each regular block is same, the relative revenue of the MP is 100%. Here, relative revenue [6] of a mining node represents the share of this node in the total mining revenue of all nodes. The blocks generated by honest miners are both stale blocks (the blocks which are not in the main chain), which means that the hash power of honest miners is wasted. Fig. 1-c illustrates the situation that the MP employs the honest mining strategy. We can see that the relative revenue of the MP is 60%. Thus, the MP can get more relative revenue

(100%, as opposed to 60% for honest mining) by using a malicious mining strategy.

However, a malicious mining strategy does not always bring more relative revenue to the MP than the honest mining strategy. See Example 2 (shown in Fig. 1-b and Fig. 1-d). In this example, the MP fails to create a main chain and its relative revenue is 0%. That is, in Example 2, the MP using a malicious strategy gains lower revenue than the honest mining strategy. From Fig. 1, we make two conclusions about malicious mining: ① The MP cannot produce more regular blocks when it adopts a malicious mining strategy than using the honest mining strategy. Hence, malicious mining can increase the relative revenue of the MP. ② When the MP adopts a malicious strategy, the growth rate of the main chain is much lower than using the honest mining strategy. Namely, malicious mining causes an increase in the regular block generation time.

In this paper, we aim to understand and quantitatively analyze the above two conclusions. We evaluate mining revenue from the perspective of relative revenue. Malicious mining not only depresses the revenue of honest miners but also creates several stale blocks, which degrades the overall performance and security of the system.

Stochastic modeling is an effective quantitative approach [12]. Researchers have conducted model-based analyses of selfish and stubborn mining attacks against blockchain systems. But they only considered attacks in Bitcoin ([6][13][14][15][18]), or only studied relative revenue of malicious miners ([3][6][16]-[20]), or ignored the impact of malicious mining on blockchain performance and security ([6][13]-[15][17]-[20]). In our work, we propose a Markov model to quantitatively evaluate the impact of selfish and stubborn mining not only on mining revenue of miners but also on the system performance and security in Bitcoin and Ethereum.

The main contributions of this work are summarized as follows.

1) We develop a novel Markov model, which can be applied to investigate eight kinds of malicious mining in Bitcoin and Ethereum. Malicious miners can employ either the honest mining strategy or a malicious mining strategy, which is the selfish mining strategy or one of the seven kinds of stubborn mining strategies. These strategies are detailed in Sec. 3. To the best of our knowledge, ours is the first model which can analyze eight kinds of malicious mining strategies in Bitcoin and Ethereum.

2) We derive the formulas to calculate relative revenue for honest miners and malicious miners in Bitcoin and Ethereum, respectively. Ethereum mining revenue comprises three kinds of rewards, but Bitcoin mining revenue comprises only one kind reward. The analysis of relative revenue has three major benefits: ① Provide quantitative relationship between the relative-revenue-optimal mining strategy for malicious miners and two miner features, including the ratio of honest miners mining in the private branch when two blocks are published simultaneously and hash power of malicious miners in Bitcoin and Ethereum, respectively. ② Provide guidelines to honest miners about setting the threshold of mining node's hash power to prevent malicious miners from profiting by selfish and stubborn mining. ③ Help honest miners design a reward mechanism to resist malicious mining.

3) We derive the formulas for calculating system performance and security metrics, including the stale block ratio, transactions per second, and the resistance against double-spending attacks. With these metrics, we can evaluate the impact of malicious mining on the blockchain system and help honest miners detect whether there are any malicious miners in the system. Our investigation of the public literature indicates that we are the first to evaluate the impact of stubborn mining on Bitcoin and Ethereum performance and security.

The rest of this paper is organized as follows. Sec. 2 presents related works. We describe the system and the developed model and metric formulas in Sec. 3. Sec. 4 shows the experiment results. Conclusions and future work are given in Sec. 5.

## 2. RELATED WORK

Since selfish mining was proposed by Eyal *et al.* [6], a number of researches were proposed to evaluate the impact of selfish mining on blockchain. Göbel *et al.* [13] analyzed the probability that honest miners mine blocks in the branch of selfish miners. Bai *et al.* [14] proposed a Markov model to evaluate the mining revenue for a selfish pool when there are two selfish pools in Bitcoin. Davidson *et al.* [15] conducted simulations to study the effect of selfish mining under different difficulty adjustment algorithms. Ritz *et al.* [16] also applied simulation to investigate the selfish mining in Ethereum. Feng *et al.* [17] proposed a two-dimensional Markov model which can describe selfish mining in Ethereum and they considered the Ethereum reward mechanism. Yang *et al.* [3] analyzed selfish mining in an imperfect Ethereum network and evaluated the impact of selfish mining on Bitcoin and Ethereum system performance and security. In this paper, when we analyze the impact of malicious mining on blockchain security, we consider a scenario that is different from [3]. Our scenario gets more accurate results, and the details are given in Sec. 3.3. All the above works only considered selfish mining. In this paper, we analyze not only selfish mining but also stubborn mining.

Stubborn mining is another kind of malicious mining, which can outperform selfish mining in some situations from the aspect of relative revenue of the MP. Stubborn mining has been studied by researchers. Nayak *et al.* [18] proposed a group of attacks, including three kinds of basic stubborn mining and four kinds of hybrid stubborn mining, which are analyzed in our paper. Wang et al. [19] analyzed the selfish mining and two kinds of stubborn mining in Ethereum. Liu et al. [20] applied Monte Carlo simulation to evaluate stubborn miners' profitability in Ethereum. In this paper, besides mining revenue, we analyze the impact of malicious mining on blockchain performance and security by modeling approach.

The fast development of blockchain technology exposes it to a range of security threats, and many researchers have been committed to building a healthy blockchain. Li et al. [21] surveyed the security risks of blockchain system and divided the risks into nine categories. Chen et al. [22] divided Ethereum system into four layers and draw insights into the vulnerability, attack, and defense of the four-layer architecture Ethereum. Ghosh et al. [23] reported various aspects related to blockchain, including taxonomy, application scenarios, vulnerabilities, attacks, and defensive methodologies. Different from them, we aim to quantitatively study malicious mining.

In the past years, except for selfish and stubborn mining attacks proposed in [6] and [18], researchers have explored other attacks against the PoW blockchain system, such as combined double-spending with Sybil attack [24], balance attack [25], block withholding attack [26], fork after withholding attack [27], and intermittent block withholding attack [11]. In addition, variants of selfish and stubborn mining attacks were developed [28]-[31]. We leave the quantitative analysis of these attacks for future work.

## 3. SYSTEM DESCRIPTION AND MODEL

This section first presents the PoW system to be studied and then the analytic model and the formulas are given. The definitions of the notations to be used are given in TABLE 1.

### 3.1 System Description

In the system, there are a large number of miners mining blocks. We assume that the total hash power of all miners is constant. The height of a block (block height) is defined as the number of regular blocks preceding this block in the blockchain. In the system, the MP can withhold a block until honest miners publish a block. For example, at a point, all miners are mining on the block with height $h-1$. The MP produces a block at height $h$ firstly but withholds it. Shortly after that, honest miners generate a block at height $h$ too and then the MP publishes their block. At this time, two blocks are at the same height $h$, and then a fork occurs, which consists of a public branch (created by honest miners) and a private branch (created by the MP).

Consensus block denotes the last block which is validated by all miners. The length of the public (private) branch is the number of blocks in the public (private) branch. Let $\Delta$ denote the length difference between the public branch and the private branch. Here $\Delta \in \{x \mid x \geq -1, x \in Z\}$. We define $\Delta = 0$ to denote that all miners mine blocks on the consensus block. $\Delta = 0'$ denotes that there is a fork with two equal-length branches, and honest miners can mine blocks on either of them. Fig. 2 gives examples of $\Delta = 2$, $\Delta = 0'$.

TABLE 1  DEFINITIONS OF NOTATIONS

| Notation | Definition |
|---|---|
| $\Delta$ | The difference in length between the public branch and the private branch. |
| $n$ | The number of branches in which honest miners can mine. $n \in \{1, 2\}$. |
| $\alpha$ | The rate of the MP generating a block. |
| $\beta$ | The rate of honest miners generating a block. |
| $\gamma$ | The ratio of honest miners that mine in the private branch when two blocks are published simultaneously. This symbol is used to denote network capacity. |
| $H_h, H_m$ | The hash power of honest miners and the MP, respectively. |
| $H_t$ | The total hash power of the system. $H_t = H_h + H_m$ |
| $S_i$ | The coefficient to distinguish different mining strategies. $i \in \{0L, 0E, 0T, L, E, T, LE\}$ |
| $p_\alpha, p_\beta$ | The probability that the malicious (honest) miners produce a block. |
| $p_{\beta p}, p_{\beta h}$ | The probabilities that honest miners produce a block in the private branch and public branch, respectively. |
| $r_j^i$ | The mining rewards when a block is produced. $i \in \{b, u, n\}$ denotes the reward type. $b$, $u$ and $n$ denote static block reward, uncle reward and nephew reward, respectively. $j \in \{h, p\}$ denotes the miner type. $h$ and $p$ denote honest miners and the MP, respectively. |
| $R_j^{BTC}, R_j^{ETH}$ | The relative revenue for the MP in Bitcoin and Ethereum, respectively. $j \in \{h, p\}$ denotes the miner type. $h$ and $p$ denote honest miners and the MP, respectively. |
| $T_i$ | The actual duration of the $i^{th}$ epoch. |
| $t_i$ | The block generation time in the $i^{th}$ epoch. |
| $b_{t,i}$ | The total number of blocks that are created in the $i^{th}$ epoch, including stale and regular blocks. |
| $b_u$ | The mean number of uncle blocks that are created in a block generation time. |
| $b_r$ | The number of regular blocks that are created in an epoch. |
| $b_p$ | The mean number of regular blocks that are created by the MP in a block generation time. |
| G1, G2 | In the scenario that there is a pool (DS pool) attacking the system by double-spending attack besides honest miners and the MP, all miners are divided into two groups. Group 1 (G1) consists of honest miners and the MP, and Group 2 (G2) consists of the DS pool. |
| $p_{G1}, p_{G2}$ | On the condition that a block is produced, $p_{G1}$ ($p_{G2}$) denotes the probability of the block produced by G1 (G2). Represent the hash power ratio of G1 and G2 in total hash power, respectively. |
| $q_{G1}, q_{G2}$ | On the condition that one of the branches length increases, the probabilities that the branch is created by G1 and G2, respectively. |

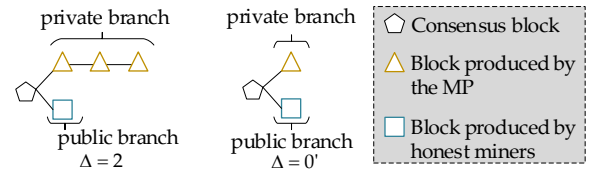

Fig. 2 Examples to illustrate $\Delta = 2$ and $\Delta = 0'$.

Before presenting the behaviors of honest miners and the MP, we first give the definitions of uncle blocks and nephew blocks, both of which are used in Ethereum. If a stale block ($B_0$) is the direct child of a regular block and it is referenced by a regular block ($B_1$), the block ($B_0$) is called an uncle block. The regular block ($B_1$) is called the nephew block. Namely, nephew block is a kind of regular block, and uncle block is a kind of stale block. In Ethereum, an uncle block can only be referenced by one regular block, and a regular block can reference up to two uncle blocks.

*1) Honest miner behaviors*

Usually, the number of honest miners is very large, and honest miners control the majority of hash power in a blockchain system. Honest miners adopt the honest mining strategy: ① An honest miner always mines blocks in the longest chain which he/she can see; ② they always publish a block as soon as the block is produced; and ③ in Ethereum, honest miners reference uncle blocks as many as they can (no more than two). Therefore, if there are two blocks (denoted as $B_A$ and $B_B$) that are published almost at the same time, a part of honest miners receive $B_A$ first and mine on it, and the rest honest miners mine blocks on block $B_B$.

*2) The malicious pool behaviors*

The MP can adopt the honest mining strategy or a malicious mining strategy (the selfish mining strategy or stubborn mining strategies). Stubborn mining is a variant of selfish mining. The selfish mining strategy was given in [3][6].

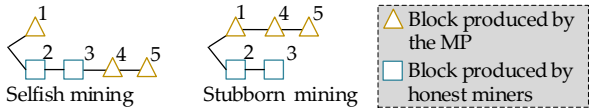

Fig. 3 An example that using a stubborn mining strategy outperforms using the selfish mining strategy.

When the MP employs the selfish mining strategy, the MP can gain an advantage at some time, which means the private branch is longer than the public branch. If the MP adopts the selfish mining strategy, the MP gives up its advantage (by publishing its unpublished blocks) or gives up the private branch (by accepting the public branch) in three cases: ① $\Delta = 2$ and honest miners produce a block, ② $\Delta = 0'$ and the MP creates a block, and ③ $\Delta = 0'$ and honest miners create a block in the public branch.

It is possible that the MP can get more relative revenue under some circumstances if it does not give up its advantage or its private branch. For example, in Fig. 3, five blocks are produced in numerical order. If the MP continues mining on the block marked '1' when the block marked '5' is produced, it gets more relative revenue (100%), as opposed to 50% in Bitcoin (and about 59% in Ethereum) for selfish mining strategy. Thus, stubborn mining is proposed. There are three basic stubborn strategies, Lead-stubborn mining (L-s), equal fork-stubborn mining (F-s), and trail stubborn mining (T-s) strategy. The following details the actions of the MP under the three basic stubborn mining strategies.

- **L-s strategy**: If $\Delta = 2$ and honest miners produce a block, the MP only publishes the first unpublished block (L-s strategy) instead of publishing the whole private branch (selfish mining strategy) (Fig. 4-a).
- **F-s strategy**: If $\Delta = 0'$ and the MP generates a block, it keeps the new block secret (F-s strategy) instead of publishing the block (selfish mining strategy) (Fig. 4-b).
- **T-s strategy**: If $\Delta = 0'$ and honest miners create a block in the public branch, the MP continues mining in the private branch (T-s strategy) instead of mining on the new block (selfish mining strategy) (Fig. 4-c). When the MP employs T-s strategy, it is possible that $\Delta = -1$. If $\Delta = -1$ and the MP generates a block, the lengths of the two branches are equal. However, at this time, all honest miners mine blocks in the public branch, because they receive the block in the public branch first, and we use $\Delta = 0''$ to denote this case.

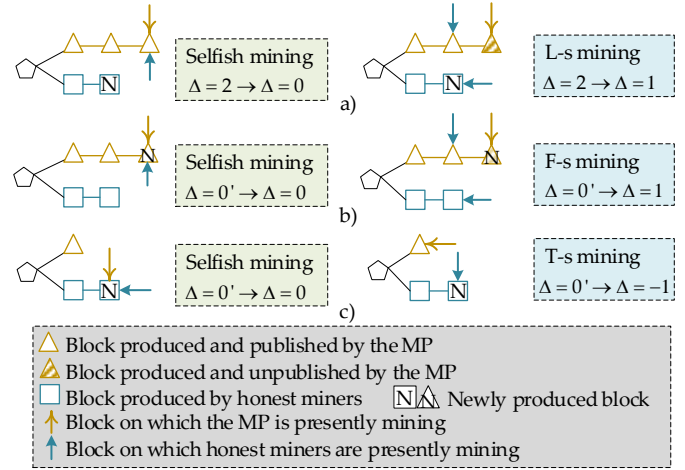

Fig. 4 The differences between each of three basic stubborn mining strategies and the selfish mining strategy.

Fig. 4 illustrates the difference between each basic stubborn mining strategy and the selfish mining strategy. The basic stubborn mining strategies are not mutually exclusive, and four hybrid stubborn mining strategies are generated by combing them. That is LF-s, LT-s, FT-s and LFT-s strategy. However, in [18], the hybrid strategies were not introduced in detail. For example, when the MP adopts FT-s strategy, $\Delta = 0''$ and the MP produces a block, it is unclear whether the new block will be published. According to [19], we define that the block will be published in this case. In the remainder of this paper, we specify multiple strategies using wildcard characters. For example, L*-s denotes L-s, LF-s, LT-s and LFT-s. In Ethereum, the MP does not reference any uncle blocks to decrease the revenue of honest miners and increases the relative revenue of the MP.

### 3.2 System Model

We use $(\Delta, n)$ to denote the system state. As the hash power of miners in Bitcoin and Ethereum are very large, the block generation process can be approximated by an exponentially distributed [3][6][14][17]. In addition, the block generation rate of miners is directly proportional to the hash power of the miners. Thus, $\alpha = H_m/H_t$ and $\beta = H_h/H_t$. There are the following two types of meaningless system states:

① State $(0',1)$. This is because that $\Delta = 0'$ means that the length of the public branch is equal to that of the private branch, and honest miners can mine blocks in public and private branch. Thus, if $\Delta = 0'$, $n$ must be equal to 2.

② State $(\Delta, 2)$ where $\Delta < 0$. If the MP adopts T-s strategy, it's possible that the private branch is shorter than the public branch by one block, and the MP still mines a block in the private branch. At this time, honest miners only mine blocks in the longest chain (the public branch). Thus, $(\Delta, 2)$ is unreasonable. Here, $\Delta < 0$.

In our model, the MP can adopt several malicious mining strategies that are described in Sec. 3.1. $S_i$ ( $i \in \{0L, 0E, 0T, L, E, T, LE\}$ ) is used to distinguish different malicious mining strategies. TABLE 2 shows the values of each $S_i$ of the eight malicious mining strategies, and 'SM' denotes selfish mining. The state transitions are illustrated in Fig. 5. The detailed state transitions are given in Appendix A of the supplementary file. $S_{LE}$ is defined to denote the MP use L*-s or F*-s strategy. Namely,

$$S_{LE} = \begin{cases} 1 & S_L = 1 \text{ or } S_E = 1 \\ 0 & \text{otherwise} \end{cases}.$$

## 3.3 Metric Formulas

This section discusses how to calculate several key metrics, including relative revenue, stale block ratio, transactions per second, and the resistance against double-spending attack. Let $p_\alpha$ and $p_\beta$ denote the probability that the MP and honest miners generate a block, respectively. $p_{\beta p}$ and $p_{\beta h}$ denote the probability that honest miners generate blocks in public branch and private branch, respectively. Note that $p_\alpha = H_m/H_t$, $p_\beta = H_h/H_t$, $p_{\beta p} = p_\beta \gamma$ and $p_{\beta h} = p_\beta (1-\gamma)$.

TABLE 2  THE MINING STRATEGIES AND THE VALUE OF $S_i$

| Strategy | $S_{0L}$ | $S_{0E}$ | $S_{0T}$ | $S_L$ | $S_E$ | $S_T$ |
|---|---|---|---|---|---|---|
| SM | 1 | 1 | 1 | 0 | 0 | 0 |
| L-s | 0 | 1 | 1 | 1 | 0 | 0 |
| F-s | 1 | 0 | 1 | 0 | 1 | 0 |
| T-s | 1 | 1 | 0 | 0 | 0 | 1 |
| LT-s | 0 | 1 | 0 | 1 | 0 | 1 |
| LF-s | 0 | 0 | 1 | 1 | 1 | 0 |
| TF-s | 1 | 0 | 0 | 0 | 1 | 1 |
| LFT-s | 0 | 0 | 0 | 1 | 1 | 1 |

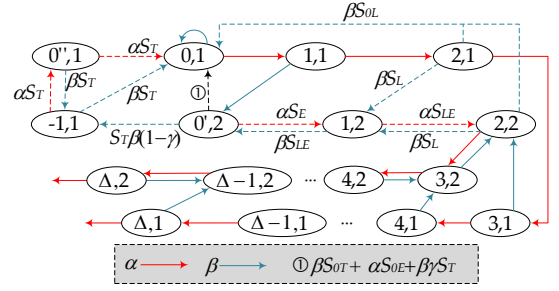

Fig. 5 The state transitions.

### 1) Relative revenue of the malicious pool

In a PoW blockchain, miners can get rewards when they produce a block. In Bitcoin system, there is only one kind mining reward, namely, static block reward. However, miners can get three kinds of mining rewards in Ethereum, including static block reward, uncle reward, and nephew reward. If the MP adopts the honest mining strategy, the relative revenue is equal to the proportion of miners' hash power of the total hash power. The mining rewards when the MP employs malicious mining strategy are given as follows. Without loss of generality, we normalize a static block reward as 1.

First, we calculate the static block reward for the MP, and the formula is given in Eq. (1).

$$\begin{aligned} r_p^b = & p_\alpha (S_{0E} + S_E p_E) \pi_{(0',2)} \\ & + p_\alpha \sum_{\Delta=1}^{\infty} \left( (S_{0L} + S_L(1 - p_{L-s}^{(\Delta+1)})) \left( \pi_{(\Delta,1)} + \pi_{(\Delta,2)} \right) \right) \\ & + p_\alpha \pi_{(0,1)} \left( p_\alpha \left( S_{0L} + S_L(1 - p_{L-s}^{(2)}) \right) + p_\beta p_{\beta p} \right. \\ & \left. + p_\alpha p_\beta (S_{0E} + S_E p_E) + S_T p_\alpha p_\beta p_{\beta h} p_{T1} \right) \\ & + S_T \left( p_\alpha \pi_{(0'',1)} + p_\alpha p_{T1} \pi_{(-1,1)} \right) \end{aligned} \quad (1)$$

where $p_{T1}$, $p_{T2}$, $p_{LE}$, $p_{L-s}^{(\Delta)}$ and $p_E$ are given in Eqs. (2) – (5), respectively.

$$p_{T1} = p_\alpha / (1 - p_\alpha p_\beta), \quad p_{T2} = S_{0T} + S_T p_\beta / (1 - p_\alpha p_\beta) \quad (2)$$

$$p_{LE} = 1 - p_\alpha p_{\beta h} \quad (3)$$

$$p_{L-s}^{(\Delta)} = p_\beta^\Delta p_{\beta h} p_{T2} / p_{LE} \quad (4)$$

$$p_E = \frac{p_\alpha \left( S_{0L} + S_L \left( 1 - p_{L-s}^{(2)} \right) \right) + p_\beta p_{\beta p} + S_T p_\alpha p_\beta p_{\beta h} p_{T1}}{1 - p_\alpha p_\beta} \quad (5)$$

Next, we compute the static block reward for honest miners (denoted as $r_h^b$).

$$\begin{aligned} r_h^b = & p_\beta \pi_{(0,1)} + S_T p_\beta \left( \pi_{(-1,1)} + p_{T2} \pi_{(0'',1)} \right) \\ & + p_\beta \left( S_{0T} + S_T ((\gamma + (1-\gamma) p_{T2})) \right) \pi_{(0',2)} \\ & + p_\beta p_{\beta h} p_{T2} \left( \pi_{(1,1)} + \pi_{(1,2)} \right) / (S_{0E} + S_E p_{LE}) \\ & + S_L p_\beta \sum_{\Delta=2}^{\infty} \left( p_{\beta h}^\Delta p_{T2} \left( \pi_{(\Delta,1)} + \pi_{(\Delta,2)} \right) / p_{LE} \right) \end{aligned} \quad (6)$$

Based on Eq. (1) and Eq. (6), we can obtain the relative revenues for honest miners and the MP in Bitcoin. They are given in Eq. (7) and Eq. (8), respectively.

$$R_h^{BTC} = r_h^b / (r_p^b + r_h^b) \tag{7}$$

$$R_p^{BTC} = r_p^b / (r_p^b + r_h^b) \tag{8}$$

However, in Ethereum, uncle reward and nephew reward are ineligible. Eq. (9) gives the uncle reward function, where $d$ is an integer and denotes the difference in block height between the uncle block and nephew block (reference distance).

$$f_u^{(d)} = \begin{cases} d/8 & d \in \{1,2,3,4,5,6\} \\ 0 & otherwise \end{cases} \tag{9}$$

The uncle reward for the MP is given in Eq. (10).

$$r_p^u = p_\alpha \left( p_\beta p_{\beta h} + S_{LE} p_\alpha p_\beta p_{\beta h}^2 / p_{LE} \right) f_u^{(1)} p_{T2} \pi_{(0,1)} \tag{10}$$
$$+ S_E p_\alpha p_\beta p_{\beta h} p_{T2} f_u^{(1)} \pi_{(0',2)} / p_{LE}$$
$$+ S_L \sum_{\Delta=1}^{\infty} \left( p_\alpha p_\beta^{\Delta+1} p_{\beta p} p_{T2} f_u^{(1)} (\pi_{(\Delta,1)} + \pi_{(\Delta,2)}) / p_{LE} \right)$$

The uncle reward for honest miners is given in Eq. (11).

$$r_h^u = p_\beta p_{\beta p} f_u^{(1)} (\pi_{(1,1)} + \pi_{(1,2)}) \tag{11}$$
$$+ \sum_{\Delta=2}^{\infty} \left( \sum_{t=0}^{\infty} \left( (p_\alpha p_{\beta h})^t \left( p_\beta + \frac{p_\alpha p_{\beta h} p_\beta p_{T2}}{S_{0E} + S_E p_{LE}} \right) f_u^{(\Delta+t)} \right) \right.$$
$$\left. \cdot p_\beta S_{0L} H^{(\Delta-2)} + p_\beta S_L \sum_{d=1}^{\infty} \left( p_{uhL}^{(\Delta,d)} p_{T2} f_u^{(d)} \right) \right) \pi_{(\Delta,1)}$$
$$+ \sum_{\Delta=2}^{\infty} \left( \sum_{t=0}^{\infty} \left( (p_\alpha p_{\beta h})^t \left( p_\beta + \frac{p_\alpha p_{\beta h} p_\beta p_{T2}}{S_{0E} + S_E p_{LE}} \right) f_u^{(\Delta+t)} \right) \right.$$
$$\left. \cdot p_{\beta p} S_{0L} H^{(\Delta-2)} + p_{\beta p} S_L \sum_{d=1}^{\infty} \left( p_{uhL}^{(\Delta,d)} p_{T2} f_u^{(d)} \right) \right) \pi_{(\Delta,2)}$$

where $p_{uhL}^{(\Delta,d)} = \begin{cases} \dfrac{(p_{\beta h})^{\Delta-1} p_{\beta p}}{p_{LE}} & , d=1 \\ \dfrac{\alpha^{\max(t-\Delta,0)} (p_{\beta h})^{\max(\Delta-1,d-1)-1} p_\beta p_{\beta p}}{p_{LE}} & , d>1 \end{cases}$

In Eq. (11), $H^{(i)} = p_{\beta h}^i + p_{\beta h}^{\max(0,i-1)} p_{\beta p}$. The detailed proofs of the correctness of Eq. (1), Eq. (6), Eq. (10), and Eq. (11) are given in proposition 1– 4 in Appendix B of the supplementary file, respectively.

The uncle reward is equal to the product of the number of uncle blocks and the corresponding uncle reward function in each case. Thus, if $f_u^{(d)} = 1$ for $d \in \{1,2,3,4,5,6\}$, the uncle reward is equal to the expected number of uncle blocks. Namely, $b_u$ is expressed as in Eq. (12) when $f_u^{(d)} = 1$ for $d \in \{1,2,3,4,5,6\}$.

$$b_u = r_p^u + r_h^u \tag{12}$$

The creator of the nephew block can get an additional reward, which is the nephew reward. As mentioned above, the MP does not reference uncle blocks and therefore $r_p^n = 0$. Thus, all uncle blocks are referenced by the blocks which are created by honest miners. A nephew reward is awarded for referencing an uncle block. Thus, the nephew reward for honest miners is given in Eq. (14).

$$r_p^n = 0 \tag{13}$$

$$r_h^n = b_u \cdot f_n^{(d)}, \text{ where } f_n^{(d)} = \begin{cases} 1/32 & d \in \{1,2,3,4,5,6\} \\ 0 & otherwise \end{cases} \tag{14}$$

With the formulas for calculating three kinds of rewards for honest miners (Eq. (6), Eq. (11), and Eq. (14)) and those for the MP (Eq. (1), Eq. (10), and Eq. (13)) in Ethereum, we can obtain the relative revenue for honest miners (in Eq. (15)) and the MP (in Eq. (16)) in Ethereum, respectively.

$$R_h^{ETH} = (r_h^b + r_h^u + r_h^n) / (r_p^b + r_p^u + r_p^n + r_h^b + r_h^u + r_h^n) \tag{15}$$

$$R_p^{ETH} = (r_p^b + r_p^u + r_p^n) / (r_p^b + r_p^u + r_p^n + r_h^b + r_h^u + r_h^n) \tag{16}$$

### 2) Stale block ratio

Stale blocks have a significant impact on blockchain difficulty adjustment, system performance and security, and therefore the stale block ratio is a key metric. The regular block creator can get static block rewards, so the regular block number is equal to the static block rewards. Assume that the expected number of blocks being produced in a block generation time is 1. Then, the mean number of regular blocks being producing in a block generation time is $r_p^b + r_h^b$. Thus, we obtain the stale block ratio, given in Eq. (17).

$$R_{stale} = 1 - (r_p^b + r_h^b) \tag{17}$$

### 3) Transactions per second

Malicious mining also has an impact on system performance. Ethereum and Bitcoin can both be payment systems, and transactions per second (TPS) is a key metric that is used to denote the system throughput. In Bitcoin and Ethereum, only the transactions in the regular blocks can be accounted, and the transactions in stale blocks are sent back to the transaction pool. Let $N_t$ and $T_{block}$ denote the number of transactions in a block and the block generation time, respectively. Then, we obtain the formula for calculating TPS for the system is given in Eq. (18).

$$TPS = (1 - R_{stale}) \cdot N_t / T_{block} \tag{18}$$

### 4) Resistance against double-spending attack

In the PoW blockchain, a large number of miners and massive hash power can protect the blockchain system against several attacks, such as double-spending attack, which is regarded as the largest security challenge of the blockchain system. However, malicious mining can reduce system security resistance capability by causing more stale blocks in the system, which wastes a lot of resources.

We consider a scenario that besides honest miners and the MP in the system that there is a pool (DS pool) launching double-spending attack, and this scenario is abbreviated as 'SDS'. There are two groups of miners in SDS. Group 1 (G1) consists of honest miners and the MP, and group 2 (G2) consists of the DS pool. The DS pool mines blocks secretly unless it attacks successfully, so G1 is not aware of G2, and G1 can be regarded as an independent system. Namely, the total hash power of G1 is $H_t$, and the hash power of honest miners and the MP is $H_h$ and $H_m$, respectively. Thus, our model and formula metrics can be used in SDS. Let $H_{ds}$ denote the hash power of the DS pool. $H_t^{SDS}$ denotes the total hash power in SDS. We can obtain,

$$H_t^{SDS} = H_t + H_{ds}, \quad p_{G1} = H_t/H_t^{SDS}, \quad p_{G2} = H_{ds}/H_t^{SDS}.$$

As the MP causes multiple stale blocks, G1 producing a block does not mean that the chain increases. $q_{G1}$ and $q_{G2}$ are given as follows.

$$q_{G1} = \frac{p_{G1}(1-R_{stale})}{p_{G1}(1-R_{stale})+p_{G2}} \qquad q_{G2} = \frac{p_{G2}}{p_{G1}(1-R_{stale})+p_{G2}}$$

To prevent double-spending attack in PoW blockchain, a transaction needs to be confirmed by multiple blocks. $N_v$ denotes the number of confirmation blocks. The formula for calculating the probability of double-spending success is given as Eq. (19).

$$P_{ds} = \begin{cases} 1 - \sum_{m=0}^{N_v} \binom{m+N_v-1}{m}\left(q_{G1}^{N_v}q_{G2}^m - q_{G1}^m q_{G2}^{N_v}\right), & \text{if } q_{G2} < q_{G1} \\ 1, & \text{if } q_{G2} \geq q_{G1} \end{cases} \quad (19)$$

A similar scenario was considered in [3]. However, in [3], the hash power of the MP which adopts the selfish mining strategy was $H_m$, and the honest miners' hash power was $H_t^{ds} - H_m = H_h + H_{ds}$. Actually, the honest miners' hash power should be $H_h$. Thus, $R_{stale}$ cannot be calculated accurately in [3].

## 4. ANALYSIS RESULTS

This section shows the analysis results. First, we use numerical analysis to study mining revenue in Sec. 4.1, the stale block ratio in Sec. 4.2, and the impacts of malicious mining on performance and security in Sec. 4.3 and Sec. 4.4. In this section, we normalize the total hash power ($H_t$) in the system as 1. The numerical analysis of our paper is conducted by using Maple 18 [32].

### 4.1 Mining Revenue Analysis

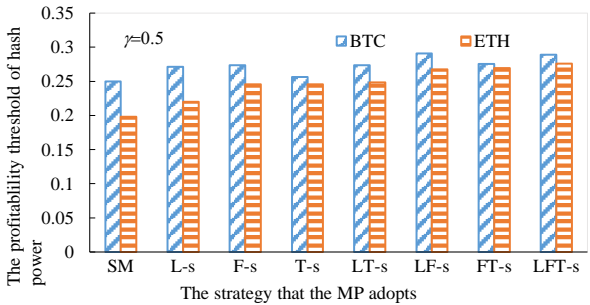

Fig. 6 The profitable threshold of the MP's hash power over strategies.

Comparing with mining honestly, the MP can get more relative revenue by malicious mining when the hash power of the MP is larger than a threshold. To avoid the MP to get unfair revenue by malicious mining, honest miners can set a threshold to limit the hash power of the MP. Thus, it is important to determine the hash power threshold. Fig. 6 shows the hash power threshold under different malicious mining strategies in Bitcoin and Ethereum (when $\gamma=0.5$). When $\gamma=0.5$, the threshold of selfish mining is 0.25, which is lower than that of the left malicious strategies. The hash power threshold of Ethereum is lower than that in Bitcoin, which means that selfish mining is easier to get unfair revenue when $\gamma=0.5$. In addition, when $\gamma=0.5$, the threshold in Bitcoin is larger than that in Ethereum no matter what the strategy is used by the MP. This is because the Ethereum reward mechanism compensates the miners who produce stale blocks, which make the MP easier to attack Ethereum than attack Bitcoin by malicious mining.

### 4.2 Stale Block Analysis

In a PoW blockchain system, the more the total hash power, the more secure and dependable the system. In our system, the stale blocks ratio is 0 when there are no malicious miners in the system. If the MP starts to attack the blockchain system by malicious mining, stale blocks are created. In Fig. 7, we show the stale block ratio when the hash power of the MP is 0.3, and the network capacity is 0.5 over different malicious mining strategies. As shown in Fig. 7, stubborn mining can cause more stale blocks than selfish mining. Namely, stubborn mining strategies have more impact on the system than selfish mining. This is because stubborn miners do not give up their private branch when selfish miners may give up. When $H_m = 0.3$ and $\gamma = 0.5$, the MP adopting stubborn mining strategies get less relative revenue than adopting the selfish mining strategy. Therefore, the MP should not employ stubborn mining strategies in this situation, because it is harmful to both honest and malicious miners.

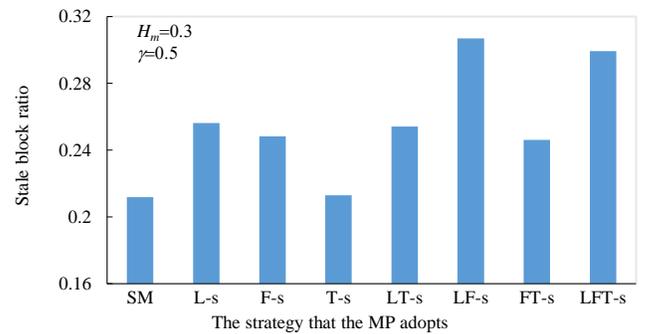

Fig. 7 Stale block ratio over the malicious strategies.

### 4.3 Blockchain Performance Analysis

Malicious miners not only get undeserved revenue, but also have a marked impact on blockchain system performance when they attack the system. If there is not malicious miner in Bitcoin system, $N_t = 2137$ and $T_{block} = 600$ [3]. We show the TPS of Bitcoin system when $\gamma = 0.5$, and the MP employs the relative-revenue-optimal strategy (the MP mines honestly when $H_m \in [0, 0.25]$ and adopts the selfish mining strategy when $H_m \in (0.25, 0.3322]$ and adopts LFT-s strategy when $H_m \in (0.3322, 0.5)$). As illustrated in Fig. 8, malicious mining can greatly decrease the growth rate of the main chain in the blockchain system, and stubborn mining leads to TPS lower than the selfish mining due to the high stale block ratio. The formulas for calculating TPS of Bitcoin is the same as that of Ethereum, and therefore, we only show TPS in Bitcoin.

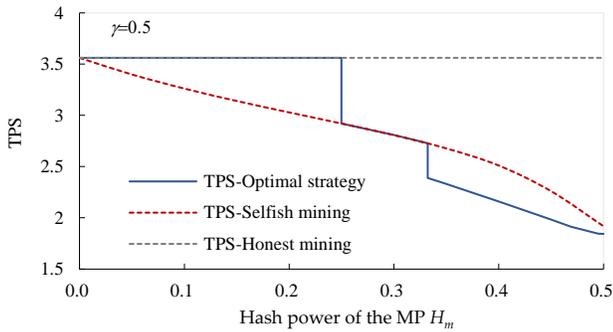

Fig. 8 Bitcoin TPS over hash power of the MP when the MP adopts the optimal mining strategy.

### 4.4 Blockchain Security Analysis

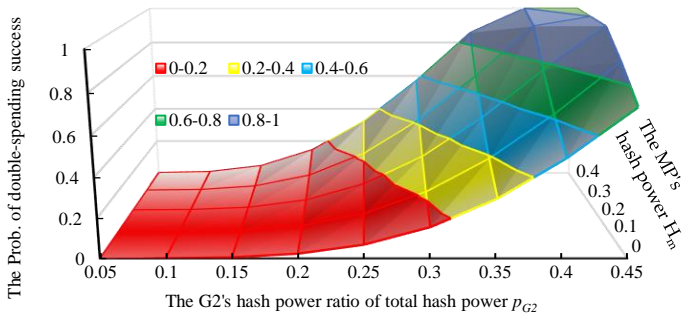

Fig. 9 Resistance to double-spending attack in Bitcoin when the MP adopts LF-s strategy.

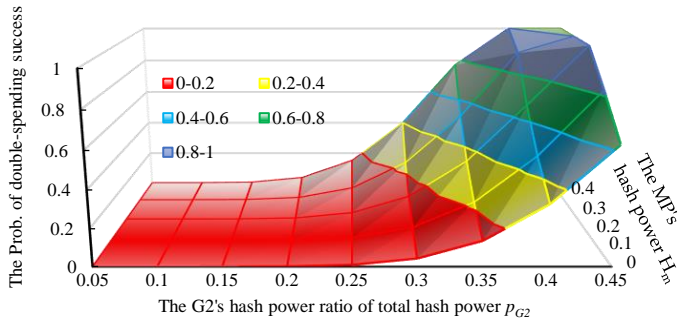

Fig. 10 Resistance to double-spending attack in Ethereum when the MP adopts LFT-s strategy.

Malicious mining also degrades the overall security of blockchain system. Due to the limited space, we do not show the impact of all malicious mining strategies on system security. We evaluate the impact of LF-s mining on Bitcoin security as an example (Fig. 9) and the impact of LFT-s mining on Ethereum security is shown in Fig. 10. The number of confirmation blocks in Bitcoin is 6, and that in Ethereum is 12 [3]. Both in Fig. 9 and Fig. 10, when we fix the hash power ratio of G2, the probability of double-spending success increases, which means that malicious mining greatly reduces the security of blockchain and makes the blockchain system more susceptible to double-spending attack.

## 5. CONCLUSIONS AND FUTURE WORK

This paper quantitatively analyzes several kinds of malicious mining strategies in Bitcoin and Ethereum systems by building a Markov model. In Bitcoin, miners can get one kind mining reward (namely, static block reward), but miners in Ethereum can earn three kinds of mining rewards (namely, static block reward, uncle reward, and nephew reward). We derive formulas of these rewards, based on which we derive formulas of computing mining relative revenue for honest and malicious miners, respectively. Moreover, the impact of malicious mining on performance and security is studied in Bitcoin and Ethereum.

The past years witnessed more than one hundred kinds of PoW blockchain based cryptocurrencies developed, including several cryptocurrencies forked from Bitcoin and Ethereum (such as Bitcoin Cash, Bitcoin Gold, Ethereum Classic) and altcoins (such as Litecoin and DASH). There are fewer miners in these cryptocurrencies, which causes that they are more susceptible to malicious mining. One of our future work direction is to apply our model and formulas to study the cryptocurrencies forked from Bitcoin and Ethereum. In addition, this paper ignores forks which are created by honest miners due to block propagation delay. However, in a realistic blockchain system, the block propagation delay does exist [33][34], especially in Ethereum network. We plan to extend our model and formulas to imperfect networks and evaluate the impact of block propagation delay on malicious mining. As mentioned in Sec. 2, besides selfish and stubborn mining, there are other attacks against block mining in PoW blockchain systems. The quantitative analysis of those attacks is also a direction of our future work.